\documentclass[twocolumn,showpacs,floatfix,aps]{revtex4}

\usepackage[dvips]{graphics}

\def \beq {\begin{equation}}
\def \eeq {\end{equation}}

\usepackage{amssymb}

\begin{document}

\title{Anisotropic singularities  in modified  gravity models}

\author{Michele Ferraz Figueir\'o}
\email{michele@fma.if.usp.br}
\affiliation{Instituto de F\'\i sica, Universidade de S\~ao Paulo,
C.P. 66318, 05315-970  S\~ao Paulo, SP,  Brazil}

\author{Alberto Saa}
\email{asaa@ime.unicamp.br}
\affiliation{
Departamento de Matem\'atica Aplicada,
IMECC--UNICAMP,
C.P. 6065, 13083-859 Campinas, SP, Brazil}


\begin{abstract}
We show that the common
 singularities present in generic modified gravity models
governed by actions of the type
$S=\int d^4x \sqrt{-g}f(R,\phi,X)$, with $X= -\frac{1}{2}g^{ab}\partial_a\phi\partial_b\phi$, are essentially
the same anisotropic instabilities associated  to
the hypersurface $F(\phi)=0$ in the case of a
non-minimal  coupling of the type  $F(\phi)R$, enlightening
the physical origin of such singularities that typically arise
in rather complex and cumbersome inhomogeneous perturbation
analyses. We show, moreover, that such  anisotropic
instabilities typically
 give rise  to dynamically unavoidable singularities, precluding
 completely
   the possibility of having physically  viable models for which
   the hypersurface $\frac{\partial f}{\partial R}=0$
   is attained.
   Some examples are explicitly discussed.
\end{abstract}

\maketitle

\section{Introduction}
In the absence of a more fundamental
 physical model based on first-principles  for
the description of the cosmic acceleration discovered more than a
decade ago\cite{acc} (see, for reviews, \cite{rev}),
many dark energy
phenomenological models have been proposed and
investigated in detail. In particular,
the questions about the stability against
 small perturbations in the initial conditions and in the model
parameters  are always the first requirement demanded to assure
the physical viability of any cosmological model. The most part
of such dark energy models belong to the general class of
cosmological models
governed by
an action  of the type (see, for instance, \cite{gen})
\beq
\label{act}
S=\int d^4x \sqrt{-g}f(R,\phi,X),
\eeq
where $R$ stands for the spacetime scalar curvature,
$\phi$ is a scalar field,  $X= -\frac{1}{2}g^{ab}\partial_a\phi\partial_b\phi$,
and
$f$ is a smooth function. Quintessence models\cite{quint},
 for instance, correspond to the choice
$
f(R,\phi,X) = \frac{1}{16\pi}R -\frac{1}{2}g^{ab}\partial_a\phi\partial_b\phi
+V(\phi).
$
Non-minimally coupled models\cite{Futamase,s2,nmc}, on the other hand, are typically of the type
\beq
\label{nmc}
f(R,\phi,X) = F(\phi)R -\frac{1}{2}g^{ab}\partial_a\phi\partial_b\phi
+V(\phi).
\eeq
Many other models discussed in the literature correspond yet
 to the case $f(R,\phi,X) = g(R,\phi) +h(\phi,X)$, including $k$-essence\cite{kess} and
the string-inspired case of a Dirac-Born-Infeld tachyonic action\cite{dbi}.
(For  more recent works, see \cite{fp}.)
The particular case of pure modified gravity
$f(R,\phi,X) = f(R)$ (see, for a recent review, \cite{fr}) has been intensively investigated as an alternative
to quintessence.
Some primordial inflationary models\cite{kinf} are also described by actions
of the type (\ref{act}).
 Since one of the proposals of
any cosmological
model  is to describe our universe without finely-tuned parameters,
a given dark energy or inflationary model would be physically viable
only if it is robust against small
perturbations in the initial conditions and in the model parameters.
This is the
question to be addressed here.

Non-minimally coupled models of the type (\ref{nmc}) are known
to be plagued by anisotropic singularities in the phase space
region corresponding to $F(\phi)=0$. For instance,
 Starobinski\cite{Starobinski} was the
first to identify the singularity corresponding to the hypersurfaces
$F(\phi)=0$, for the case of conformally coupled anisotropic solutions.
Futamase and co-workers\cite{Futamase} identified the same
kind of singularity
in the context of chaotic inflation in $F(\phi)=1-\xi\phi^2$ theories
(See also \cite{s2}). In \cite{sing}, it is shown that   such
kind of singularities are generically
 related to anisotropic instabilities.

Many authors have described different singularities corresponding to
$\frac{\partial f}{\partial R}=0$ in general models like (\ref{act}) (see,
for instance, \cite{gen}) or,
more commonly, in pure $f(R)$ gravity models (see, for instance,
\cite{fr,singfr}). Such singularities
appear typically in rather complex and cumbersome
inhomogeneous perturbation analyses, obscuring their physical
origin and cause.
In this work, we show that these singularities are essentially
due to anisotropic instabilities, in a similar way to those ones
described in \cite{sing} for models of the type (\ref{nmc}). Moreover,
we show that such instabilities typically
 give rise  to dynamically unavoidable singularities, rendering
 the original model physically unviable.

One can advance that there are some geometrically
special regions on the phase space of the
model in question by an elementary
  analysis of the equations derived from the
action (\ref{act}).
They are the generalized Klein-Gordon equation
\beq
\label{kg}
D_a\left(f_{,X} \partial^a\phi \right) + f_{,\phi} = 0,
\eeq
and the Einstein equations
\begin{eqnarray}
\label{ee}
FG_{ab} &=& \frac{1}{2}\left(f-RF\right)g_{ab} + D_aD_b F
- g_{ab}\Box F \nonumber \\ &-& \frac{1}{2} f_{,X} \partial_a\phi  \partial_b\phi,
\end{eqnarray}
where $F =F(R,\phi,X)\equiv\frac{\partial f}{\partial R}$.
We will consider here the simplest anisotropic homogeneous cosmological
model, the Bianchi type I, whose spatially flat metric is given by
\beq
\label{metric}
ds^2 = -dt^2 + a^2(t)dx^2 + b^2(t)dy^2 + c^2(t)dz^2.
\eeq
The dynamically relevant quantities in this case are
\beq
H_1 = \frac{\dot{a}}{a}, \quad H_2 = \frac{\dot{b}}{b}, \quad
{\rm and\ } H_3 = \frac{\dot{c}}{c}\ .
\eeq
For such a metric and with a homogeneous scalar field $\phi=\phi(t)$,
Einstein Eq. (\ref{ee}) can be written as
\begin{eqnarray}
\label{ec}
FG_{00} &=& -\frac{1}{2}\left(f-FR \right) - (H_1+H_2+H_3)\dot{F} \nonumber \\
&& +
\frac{1}{2}f_{,X}\dot{\phi}^2,\\
\label{e1}
\frac{FG_{11}}{a^2} &=& \frac{1}{2}\left(f-FR \right) + (H_2+H_3)\dot{F}   +\ddot{F} ,
\\\label{e2}
\frac{FG_{22}}{b^2} &=& \frac{1}{2}\left(f-FR \right) + (H_1+H_3)\dot{F}   +\ddot{F} ,
\\\label{e3}
\frac{FG_{33}}{c^2} &=& \frac{1}{2}\left(f-FR \right) + (H_1+H_2)\dot{F}   +\ddot{F} ,
\end{eqnarray}
and the generalized Klein-Gordon equation will read
\beq
\label{kgg}
\frac{d}{dt}\left(f_{,X}\dot{\phi} \right) +
(H_1+H_2+H_3)f_{,X}\dot{\phi} - f_{,\phi} = 0.
\eeq
Notice that (\ref{kgg}) is a second order differential equation for $\phi$,
while Eqs. (\ref{e1})-(\ref{e3}) form a higher order system of ordinary differential equations. Since $F =F(R,\phi,X)$, the term corresponding to
$\ddot{F}$ involves, in fact, second derivatives of $R$ and, consequently, third
derivatives of $H_i$, $i=1,2,3$. Thus, the corresponding phase space $\cal M$ is 11-dimensional
and spanned by the variables $(\phi,\dot{\phi},H_1,\dot{H}_1,\ddot{H}_1,H_2,\dot{H}_2,\ddot{H}_2,H_3,\dot{H}_3,\ddot{H}_3)$.
Eq. (\ref{ec}) corresponds to the energy constraint. It restricts  the
solutions of (\ref{e1})-(\ref{kgg}) on a certain (vanishing energy)
hypersurface $\cal E$ of $\cal M$. Thus, effectively, the solutions of
(\ref{e1})-(\ref{kgg}) are constrained to the 10-dimensional manifold
${\cal E}\in \cal M$.

It is quite simple to show that Eqs. (\ref{e1})-(\ref{e3}) are
not compatible, in general,
 on the hypersurface $\cal F$  of  $\cal M$
 corresponding to the region where $F(R,\phi,X)=0$.  Subtracting
(\ref{e2}) and (\ref{e3}) from (\ref{e1}) we have,   respectively, on
such hypersurface
\beq
\label{cau}
 (H_1-H_2)\dot{F} = 0,\ {\rm and\quad }
(H_1-H_3)\dot{F} = 0.
\eeq
Hence, Eqs. (\ref{e1})-(\ref{e3})
  cannot be fulfilled in general for anisotropic metrics. As it
will be shown, the hypersurface $\cal F$ indeed corresponds
a  geometrical
singularity for anisotropic spacetimes which cannot be dynamically
 prevented in general
by requiring, for instance, that $\dot{F}=0$
  on the hypersurface $\cal F$ as suggested naively from (\ref{cau}).
Furthermore,  the Cauchy problem for the Eqs. (\ref{e1})-(\ref{kgg})
is ill-posed on this hypersurface, since one cannot
choose general initial conditions on it.

\section{The singularity}

In order to study the geometrical nature of  the singular
hypersurface $\cal F$, let
us consider the Einstein Eqs. (\ref{ec})-(\ref{e3}) in
detail. For the metric (\ref{metric}), we have the following
identities
\begin{eqnarray}
%\label{gg}
G_{00} &=& H_1H_2 + H_2H_3 + H_1H_3,   \\
G_{11} &=& a^2\left( \dot{H}_1 + H_1(H_1+H_2+H_3) - \frac{1}{2}R\right),
 \\
G_{22} &=& b^2\left( \dot{H}_2 + H_2(H_1+H_2+H_3) - \frac{1}{2}R\right), \\
G_{33} &=& c^2\left( \dot{H}_3 + H_3(H_1+H_2+H_3) - \frac{1}{2}R\right),
 \\
R &=& 2( \dot{H}_1 + \dot{H}_2  + \dot{H}_3
+ {H}_1^2 + {H}_2^2 + {H}_3^2  \nonumber \\
 && + H_1H_2 + H_2H_3 + H_1H_3).
\end{eqnarray}
Now, we introduce  the
new dynamical variables $p=H_1+H_2+H_3$, $q=H_1-H_2$, and
$r=H_1-H_3$. Notice that
\beq
\label{R}
R = 2\dot{p} + \frac{2}{3}\left(2p^2 + q^2 + r^2 - qr \right),
\eeq
implying that $\ddot{R}$ involves terms up to third order derivative
in $p$ and up to second order in $q$ and $r$.
In terms of the new dynamical variables, Einstein Eqs. (\ref{e1})-(\ref{e3})
  can be cast in the form
\begin{eqnarray}
\label{ep}
3\ddot{F} &=& \left(\dot{p} + p^2 \right)F - \frac{3}{2} f -2 p\dot{F}, \\
\label{eq}
q\dot{F} &=&  - \left(\dot{q} + qp \right)F, \\
\label{er}
r\dot{F} &=&  - \left(\dot{r} + rp \right)F.
\end{eqnarray}
As to the energy constraint (\ref{ec}), we have
\begin{eqnarray}
\label{ec1}
 \frac{1}{3}\left(
p^2 + qr - q^2 - r^2
\right)F  &+& p\dot{F}      \nonumber  \\
+ \frac{1}{2}\left(f-FR \right) &=& \frac{1}{2}f_{,X}\dot{\phi}^2 ,
\end{eqnarray}
and the generalized
Klein-Gordon equation (\ref{kg}) reads simply
\beq
\label{kg1}
\left(f_{,X} + f_{,XX} \dot{\phi}^2 \right)\ddot{\phi}   +
\left(f_{,XR}\dot{R} + f_{,X\phi}\dot{\phi}+
pf_{,X} \right)\dot{\phi} - f_{,\phi} = 0.
\eeq
Notice that the Eqs. (\ref{ep})-(\ref{kg1}) do not involve
the terms $\dddot{q}$ and $\dddot{r}$. Moreover, Eqs. (\ref{eq}) and
(\ref{er}) are, respectively, first order differential equations for
$q$ and $r$,  from which the terms involving first and second derivative of
$q$ and $r$ present in the terms $\dot{F}$ and $\ddot{F}$ of (\ref{ep})
and (\ref{ec1})  can be
evaluated directly. The order reduction of the system of differential
equations attained with the introduction
of the new dynamical variables implies that the phase space $\cal M$
is not 11, but 7-dimensional and spanned by the variables
$(\phi,\dot{\phi},p,\dot{p},\ddot{p},q,r)$. The solutions are still
constrained to the hypersurface ${\cal E}\in \cal M$ corresponding
to the energy constraint (\ref{ec1}). It is clear, however, that the
manifold ${\cal E}$ is, in fact, 6-dimensional.

There is still a further dynamical restriction on the solutions
of (\ref{ep})-(\ref{kg1}). From
  (\ref{eq}) and (\ref{er}), one has
 \beq
 \label{restr}
 r\dot{q}-q\dot{r}=0,
 \eeq
  implying that $q(t)/r(t)$ is a constant
of motion
fixed only by the initial conditions.
Suppose the initial ratio is $q(0)/r(0)=\gamma$: this would imply that
$(H_1-H_2)=\gamma(H_1-H_3)$ for all $t$, leading to, for instance,
$c^\gamma(t) \propto a^{\gamma-1}(t)b(t)$ in the
metric (\ref{metric}). This simplification
is a consequence of the  scalar character of our homogeneous
source field,
and it is also present\cite{sing} in  the non-minimally coupled
case given by actions of the form (\ref{nmc}). Let $\cal Q$ be
the hypersurface corresponding to $q/r$ constant. Finally,
the solutions
of (\ref{ep})-(\ref{kg1}) are necessary restricted to the
5-dimensional submanifold ${\cal Q}\cap{\cal E}$ of $\cal  M$.

A closer analysis of Eqs. (\ref{eq}) and (\ref{er}) reveals the presence of the
singularity.
They can be written as
\begin{eqnarray}
\label{eq1}
 \dot{q} &=&  - \left(p + \frac{\dot{F}}{F}\right)q, \\
\label{er1}
 \dot{r} &=&  - \left(p + \frac{\dot{F}}{F}\right)r.
\end{eqnarray}
 In general, the right-hand side of these equations
diverge on the hypersurface $\cal F$ corresponding to
$F(R,\phi,X)=0$, unless $q=r=0$. The first observation
is that (\ref{eq1}) and (\ref{er1}) imply, in general, that, if
$q$ (or $r$) vanishes for some $t$, it will
 vanish for
any $t$.
This is why such kind of singularity can be
evaded in homogeneous and isotropic situations.
We will return to this point in the next section, with
an explicit example.
For any
physically viable cosmological model,   small amounts  of
anisotropy, corresponding to small $q$ and $r$, must stay
bounded during the cosmological history. In fact, it is
desirable that they diminish, tending towards an isotropic
situation. However, this does not happen in general if
 $F(R,\phi,X)=0$ in (\ref{eq1}) and (\ref{er1}). Let us
 assume that $\dot{F}\ne 0$ on the hypersurface $\cal F$.
 (We will return to this point latter.) In this case, if
 any anisotropic solution crosses $\cal F$, necessarily
 $\dot{q}$ and $\dot{r}$ will diverge, corresponding to
 a real spacetime geometrical singularity,
as one can check
by considering the Kretschman invariant $I=R_{abcd}R^{abcd}$, which
for the metric (\ref{metric}) is given by
\begin{eqnarray}
\label{Kret}
\frac{1}{4}I &=&
\left(\dot{H}_1+H_1^2\right)^2 +
\left(\dot{H}_2+H_2^2\right)^2 +
\left(\dot{H}_3+H_3^2\right)^2   \nonumber \\
&+&   H_1^2H_2^2 + H_1^2H_3^2 + H_2^2H_3^2.
\end{eqnarray}
As one can see, $I$ is the sum of non negative terms.
Moreover, any divergence of the variables $H_1$, $H_2$, $H_3$, or of
their time derivatives, would suppose a divergence in $I$,
characterizing a real geometrical singularity. Since
the relation between the variables $p$, $q$, $r$, and
$H_1$, $H_2$, $H_3$ is linear, any divergence of the
first, or of their time derivative, will suppose a divergence
in $I$.

There are two basically distinct situations where the singularity
corresponding to the hypersurface $\cal F$ could be evaded dynamically.
We will show that both are very unlike to occur
  in physical situations. The first one corresponds to the case when
  the hypersurface $\cal F$ belongs to some dynamically unaccessible region.
  In such a case we, of course, do not face any singularity, since
$F(R,\phi,X)$ will never vanish along a solution of the system.
This would  be equivalent to state that ${\cal F}\cap{\cal E}\cap{\cal Q} = \emptyset$.
In our case, it would imply, from (\ref{ec1}), that the equation
\beq
\label{conc}
 p\dot{F}
= Xf_{,X}   -   \frac{1}{2} f
\eeq
has no solution in $\cal M$. This would correspond to a quite
concocted and artificial function $f$.
In particular, for all models we could find in the literature
 having the  hypersurface $\cal F$, the equation (\ref{conc}) has
 solutions.

The second situation corresponds to the already mentioned
 case where $\dot{F}=0$
on the hypersurface $\cal F$.
From (\ref{conc}), we see that this requires necessarily
 that the function $f$
be homogeneous of degree
 $\frac12$ in the variable $X$ on $\cal F$. Again, a highly artificial
 situation.

Any point on the energy constraint hypersurface ${\cal E}$ is, in principle,
a dynamically possible point. Moreover, it is desirable for any cosmological
model free of finely-tuned parameters that any point or, at least, a large
 region  of  ${\cal E}$ could be chosen as the initial condition for a
 cosmological evolution. This, of course, includes also the neighborhood of
 the hypersurface ${\cal F}$  provided that ${\cal F}\cap {\cal E}\ne \emptyset$.

\section{An explicit example}

The singularities described in the precedent section imply that
any   model governed by an action of the type (\ref{act}) having
a hypersurface $\cal F$ will certainly present severe  anisotropic
instabilities that will render it physically unviable. Let us
work out an explicit example in order to illustrate the dynamical
role of
such anisotropic instabilities.  The pure modified gravity model
\beq
\label{frc}
f(R) = R - \alpha R_*\ln\left(1+\frac{R}{R_*} \right),
\eeq
where $\alpha$ and $R_*$ are free positive parameters,
was recently proposed\cite{frc} as a viable model to describe the
recent cosmic acceleration. Such a model has a hypersurface $\cal F$ corresponding
to $f'(R)=F(R)=0$, where
\beq
F(R)= 1 - \frac{\alpha R_*}{R+R_*}.
\eeq
 In  \cite{frc}, it is assumed a universe filled with radiation
 and dark matter, but,
for our purposes here, it is enough to consider the pure geometrical
Lagrangian given by (\ref{frc}).

Let us start, as in \cite{frc},  assuming a
homogeneous and isotropic universe $H_1=H_2=H_3=H$. Einstein
Eqs. (\ref{ec})-(\ref{e3}) for this case would correspond simply
to the energy constraint
\beq
\label{ec2}
6H\dot{R}F'(R) + f(R) - RF(R) + 6H^2F(R)= 0,
\eeq
and to the generalized Friedman equation
\begin{eqnarray}
\label{ee2}
  \ddot{R}F'(R) &+ &\left(2HF'(R)  +  \dot{R}F''(R) \right)\dot{R}   \\
&+& \frac{1}{2}f(R)- \left(\dot{H} + 3H^2 \right)F(R) = 0, \nonumber
\end{eqnarray}
where $R=6\dot{H}+12H^2$ in this homogeneous and isotropic case. Note that
\beq
\label{FP}
F'(R) = \frac{\alpha R_*}{\left(R+R_* \right)^2} > 0,
\eeq
for $R+R_*\ne 0$.
Eq. (\ref{ee2}) is a third order differential equation for $H$. Hence,
 the relevant
phase space is 3-dimensional and spanned by the variables
$(H,\dot{H},\ddot{H})$, but the solutions are in fact
constrained to the
2-dimensional manifold $\cal E$ corresponding to the energy constraint
(\ref{ec2}). The manifold $\cal E$ is an ordinary smooth surface,
 with a single
value of $\ddot{H}$ assigned to each pair $(H,\dot{H})$, provided
$H\ne 0$ and $R+R_*\ne 0$. Thus, the solutions of (\ref{ee2})
can be conveniently projected
on the plane  $(H,\dot{H})$, without any loss of dynamical information.

\begin{figure}[ht]
\resizebox{1\linewidth}{!}{ \includegraphics*{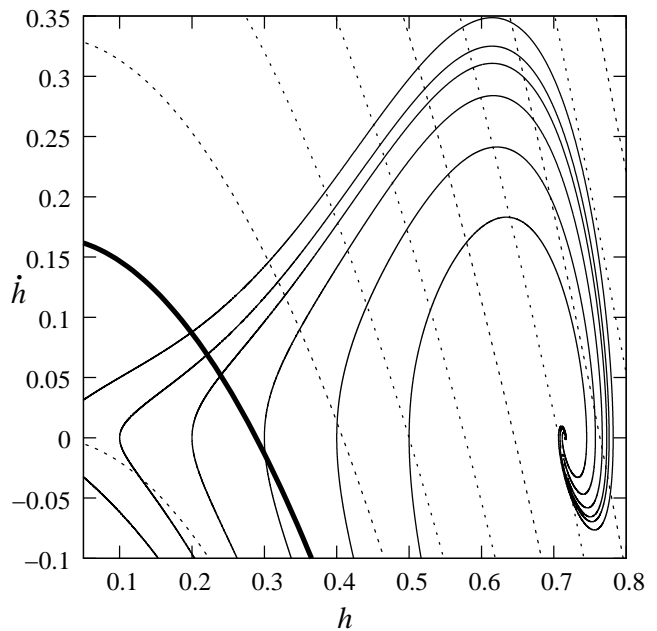}}
\caption{Typical   trajectories for the model (\ref{frc})
projected on the ($h,\dot{h}$) plane,
  with $\alpha =2$ (the same value adopted in the analysis of
  \cite{frc}).
The (attractive de Sitter) fixed point in this case corresponds to
$h\approx 0.7157$. The $\rho$-constant parabolas are shown. The solid one,
in particular, correspond to $\rho=\alpha-1$ (the singular ${\cal F}$ surface,
defined by $F(R)=0)$. The regions corresponding to $F(R)>0$ and $F(R)<0$
are, respectively, the region above and the region below the parabola $\rho=\alpha-1$.
 As one can see, homogeneous and isotropic solutions
can cross without problems the singular surface. }
\label{fig}
\end{figure}

It is convenient to work with the dimensionless
quantities $H=\sqrt{R_*}h$, $\sqrt{R_*}\tau = t$, $R=\rho R_*$.
The phase space for this model is quite simple. There are only
 two
 fixed points corresponding to $h=\pm\sqrt{\tilde{\rho}/12}$, where
 $\tilde{\rho}$ is the positive solution of the equation
 \beq
 2\alpha \ln(1+\rho) = \rho + \alpha\frac{\rho}{1+\rho}.
 \eeq
 This solution
  exists and is unique provided that $\alpha > 1$.
  Both equations (\ref{ec2}) and (\ref{ee2}) are invariant under
  the transformation $\tau \rightarrow -\tau$
  and $h \rightarrow -h$, implying that the $h$ negative  portion of
  the phase space can be obtained from the positive one by means of
  a time
  reversal operation. Typical trajectories projected on the $(h,\dot{h})$
  plane of the phase space are depicted in Fig. \ref{fig}.
Notice that the surfaces corresponding to $\rho = R/R_*$ constant are simple parabolas
$6\dot{h} + 12h^2 = \rho$ in the  $(h,\dot{h})$ plane. For the model in
question, the ${\cal F}$ surface corresponds to one of these parabolas,
 namely $\rho=\alpha -1$.
The point here is that the existence of such a surface
does not imply any
singular behavior for the equation (\ref{ee2}). For instance,
the solutions crossing
the surface $F(R)=0$ depicted in \ref{fig} are perfectly regular.
 Hence, homogeneous
and isotropic solutions can cross without problem the singular
hypersurface.

Suppose now the system has a small amount of anisotropy, {\em i.e.},
$|q|  \ll |p| $ and $|r|  \ll |p| $. In this case, we have from (\ref{R})
$R \approx 2\dot{p} + \frac{4}{3}p^2$
 and the equation  (\ref{ep})  for $p$
 will be essentially the same (\ref{ee2}) obeyed by $H$  in
 the isotropic case, provided the anisotropy is indeed kept
 small along the solutions. For the amounts of anisotropy $q$ and
 $r$, however, the relevant
equations will be (\ref{eq1}) and (\ref{er1}).
Since we know from (\ref{restr}) that $r(t) = \gamma q(t)$, we
can  consider here  only the variable $q$
\beq
\label{eq3}
\dot{q} = -\left(p + \frac{F'(R)}{F(R)}\dot{R} \right)q.
\eeq
In any region of the phase space far from the surface $F(R)=0$,
the right-handed side of (\ref{eq3}) is well behaved. Moreover,
from the energy constraint (\ref{ec1}), we have
\beq
\label{ec3}
p+\frac{F'(R)}{F(R)}\dot{R} = \frac{2}{3}p
+ \frac{\gamma^2 - \gamma +1}{3}\frac{q^2}{p}
+ \frac{1}{2p}\left(R-\frac{f(R)}{F(R)} \right).
\eeq
A closer analysis reveals that
\beq
RF(R) - f(R) = \alpha R_*\left(\ln(1+\rho) - \frac{\rho}{1+\rho} \right)\ge 0,
\eeq
with the equality holding only for $\rho=0$, implying that
for the region $F(R) > 0$,   at
least,  the quantity between parenthesis in (\ref{eq3})
is positive, leading indeed to an isotropization of the solutions.
For   regions close to the surface $F(R)=0$, on the other hand,
the situation is qualitatively different.
From (\ref{ec3}), we have that the
 right-handed side of (\ref{eq3})  diverges on
  the surface $F(R)=0$.
If an anisotropic solution reaches  such
surface, we have from
Eq. (\ref{eq3})  that $\dot{q}$ diverges,
implying that this model does not admit any amount of anisotropy at all,
precluding any possibility of constructing a realistic model based solely
in the geometric Lagrangian (\ref{frc}).
Similar results hold also for the other functions $f(R)$ discussed in
\cite{frc}, namely
\beq
f(R) = R  - \alpha R_*\left(1 + \frac{R}{R_*} \right)^\beta,
\eeq
with $\beta\in (0,1)$.

\section{Final Remarks}

The singularities associated with the hypersurface $F(R,\phi,X)=0$
described here are not new. They have been discovered
and rediscovered many times for many
different models in rather complex and cumbersome inhomogeneous
perturbation analysis around a given well behaved background solution.
Our results, however, enlighten the physical origin of such
singularities. They arise already in the background level and are
related to anisotropic expansion rates. Any solution crossing
the hypersurface $F(R,\phi,X)=0$ will not admit, in general, any
amount of anisotropy, otherwise it will certainly develop a catastrophic
geometrical singularity with, for instance,
the blowing up of the Kretschman invariant
(\ref{Kret}). This, in fact, precludes the possibility of constructing
a realistic model with solutions crossing the hypersurface $F(R,\phi,X)=0$
since we would have qualitatively distinct behavior for arbitrarily
close homogeneous solutions: a perfect isotropic  and a slightly anisotropic one.

\section*{Acknowledgements}

The authors
wish to thank Prof. V. Mukhanov and his group
for the warm hospitality at the
Arnold Sommerfeld Center for Theoretical Physics of the
Ludwig-Maximilians University of Munich, Germany, where this work was
carried out.
This work was supported by FAPESP (Brazil), DAAD (Germany)
and   CNPq (Brazil).

\end{document}